%% file: dis_procs-3.tex
\documentclass[a4paper]{jpconf}

\usepackage{amsmath,amssymb}
\usepackage{lipsum}
\usepackage{enumitem}
\usepackage{soul}
\usepackage{graphicx}
\usepackage{color}
\usepackage{slashed}
\usepackage{bbold}
\usepackage{wasysym}
\usepackage{graphicx}
\usepackage{cite}

\usepackage{multirow}

\newcommand{\TeV}{{\ensuremath\rm TeV}}
\newcommand{\GeV}{{\ensuremath\rm GeV}}

\newcommand{\fb}{\ensuremath\rm fb}
\newcommand{\pb}{\ensuremath\rm pb}
\newcommand{\ab}{\ensuremath\rm ab}

\newcommand{\mgfive}{\texttt{MG5\_aMC@NLO}}
\newcommand{\checkmate}{\texttt{CheckMATE}}

\newcommand{\HSv}[1]{\texttt{HiggsSignals-#1}}
\newcommand{\HBv}[1]{\texttt{HiggsBounds-#1}}

\newcommand{\lam}{\lambda}

\begin{document}

\title{The Inert Doublet Model at current and future colliders}

\author{J Kalinowski$^1$, W Kotlarski$^2$, T Robens$^3$, D Sokolowska$^4$, A F \.Zarnecki$^1$}
\address{$^1$ Faculty of Physics, University of Warsaw,ul.~Pasteura 5, 02--093 Warsaw, Poland}
\address{$^2$ Institut f\"ur Kern- und Teilchenphysik, TU Dresden, 01069 Dresden, Germany}
\address{$^3$ Theoretical Physics Division, Rudjer Boskovic Institute, 10002 Zagreb, Croatia}
\address{$^4$ International Institute of Physics, Universidade Federal do Rio Grande do Norte, Campus Universitario, Lagoa Nova, Natal-RN 59078-970, Brazil}
\ead{Jan.Kalinowski@fuw.edu.pl, Wojciech.Kotlarski@tu-dresden.de, dsokolowska@iip.ufrn.br, trobens@irb.hr, Filip.Zarnecki@fuw.edu.pl}

\begin{abstract}

We discuss the status of the Inert Doublet Model, a two-Higgs doublet model that obeys a discrete $Z_2$ symmetry and provides a dark matter candidate. We discuss all current theoretical and experimental constraints on the model as well as discovery prospects at current and future colliders.
\end{abstract}

\section{Introduction}
The Inert Doublet Model (IDM) \cite{Deshpande:1977rw,Cao:2007rm,Barbieri:2006dq} is an intriguing extension of the Standard Model (SM) scalar sector which features a dark matter candidate. It is a two Higgs doublet model with the scalar potential
\begin{equation*}\begin{array}{c}
V=-\frac{1}{2}[m_{11}^2(\phi_S^\dagger\phi_S)\!+\! m_{22}^2(\phi_D^\dagger\phi_D)]+
\frac{\lambda_1}{2}(\phi_S^\dagger\phi_S)^2\! 
+\!\frac{\lambda_2}{2}(\phi_D^\dagger\phi_D)^2 +\!\lambda_3(\phi_S^\dagger\phi_S)(\phi_D^\dagger\phi_D)\! \\
\!+\!\lambda_4(\phi_S^\dagger\phi_D)(\phi_D^\dagger\phi_S) +\frac{\lambda_5}{2}[(\phi_S^\dagger\phi_D)^2\!
+\!(\phi_D^\dagger\phi_S)^2]
\end{array}\label{pot}\end{equation*}
obeying an additional discrete $Z_2$ symmetry (called $D$-symmetry) under which
%\begin{equation}
$\phi_S\to \phi_S, \,\, \phi_D \to - \phi_D, \,\,
\text{SM} \to \text{SM}. $
%\end{equation}
Exact $D$-symmetry implies that only  $\phi_S$ can acquire {a} nonzero vacuum expectation value ($v$) and the $\phi_S$ doublet plays the same role as the corresponding  doublet in the SM, {providing} the SM-like Higgs particle. {The same} $Z_2$ symmetry implies that 
the second doublet, the inert (or dark) $\phi_D$, does not mix with the SM-like field {from} $\phi_S$ and  does not couple to the SM matter fields. 
The dark sector  {contains} four new particles:  two charged and two neutral ones, labelled $H^\pm$ and $H,A$, respectively. The lightest one is therefore stable and 
we choose {$H$ as the} DM candidate. {The IDM has been widely explored in the literature}; cf. e.g. \cite{Gustafsson:2012aj,Arhrib:2012ia,Swiezewska:2012eh,Goudelis:2013uca,Krawczyk:2013jta,Aoki:2013lhm,Ho:2013spa,Arhrib:2013ela,Ginzburg:2014ora,Belanger:2015kga,Blinov:2015qva,Ilnicka:2015jba,Hashemi:2015swh,Diaz:2015pyv,Arhrib:2015hoa,Poulose:2016lvz,Datta:2016nfz,Kanemura:2016sos,Akeroyd:2016ymd,Belyaev:2016lok,Eiteneuer:2017hoh,Wan:2018eaz,Ilnicka:2018def,Belyaev:2018ext,Kalinowski:2018kdn,Kalinowski:2018ylg,Dercks:2018wch} for recent discussions. \\
After electroweak symmetry breaking the model contains seven free parameters. Agreement with the Higgs boson discovery and electroweak precision observables fixes the SM-like Higgs mass  {$m_h$} and $v$,  {and we are left with} five free parameters:
%\begin{\eqn}\label{eq:physbas}
the dark scalar masses $m_H, m_A, m_{H^{\pm}}$ and two couplings $\lam_2, \lam_{345}\,\equiv\,\lam_3+\lam_4+\lam_5$.
%\end{\eqn}

\section{Experimental and theoretical constraints \label{sec:constraints}}

Here we list the constraints applied in our studies ({see} refs \cite{Krawczyk:2013jta,Ilnicka:2018def,Kalinowski:2018ylg} for more details):\\[1mm]
{\it Theoretical constraints:}  included are positivity of the potential \cite{Nie:1998yn}, condition to be in the inert vacuum \cite{Ginzburg:2010wa}\footnote{{For the recast studies in \cite{Dercks:2018wch}, that condition was relaxed. See that reference for a detailed discussion.}} and perturbative unitarity \cite{Chanowitz:1985hj,Ginzburg:2005dt}. Some of these constraints are tested via the publicly available two-higgs doublet model calculator (\texttt{2HDMC}) tool \cite{Eriksson:2009ws}.\\[1mm]
{\it Collider constraints:}  included are agreement with electroweak precision tests \cite{Baak:2014ora} via oblique parameters \cite{Altarelli:1990zd,Peskin:1990zt,Peskin:1991sw,Maksymyk:1993zm}, agreement with electroweak gauge boson widths \cite{PhysRevD.98.030001}, requirement of a short-lived charged scalar\footnote{We here take a rough estimate on the charged scalars lifetime; see e.g. \cite{Heisig:2018kfq} for a recent recast of LHC searches.}, agreement with recasts of LEP and LHC searches \cite{Lundstrom:2008ai,Belanger:2015kga,Dercks:2018wch}, the total width \cite{Sirunyan:2019twz}, the invisible branching ratio of the 125 \GeV~Higgs \cite{Khachatryan:2016whc} and the branching ratio $h\,\rightarrow\,\gamma\gamma$ \cite{Khachatryan:2016vau}, agreement with current collider measurements of the Higgs signal strength as well as null-results for additional scalar searches at the LHC, where we make use of {\HBv5.4.0beta} \cite{Bechtle:2008jh, Bechtle:2011sb, Bechtle:2013wla,Bechtle:2015pma} and {\HSv2.2.3beta}~\cite{Bechtle:2013xfa}.\\[1mm]
{\it Astrophysical constraints:} included are agreement with results from direct detection experiments \cite{Aprile:2018dbl} and  that the dark matter relic density for our model does not lead to overclosing of the universe; i.e., we require the relic density to be at most within a two sigma range of the last value measured by the Planck experiment, i.e. $\Omega_c\,h^2\,\leq\,0.1224$ \cite{Aghanim:2018eyx}. Dark matter predictions have been obtained using \texttt{micrOmegas} version 4.3.5 \cite{Barducci:2016pcb}.
\vspace{3mm}

\begin{center}
\begin{figure}
\begin{center}
\begin{minipage}{0.45\textwidth}
\begin{center}
\includegraphics[width=0.9\textwidth]{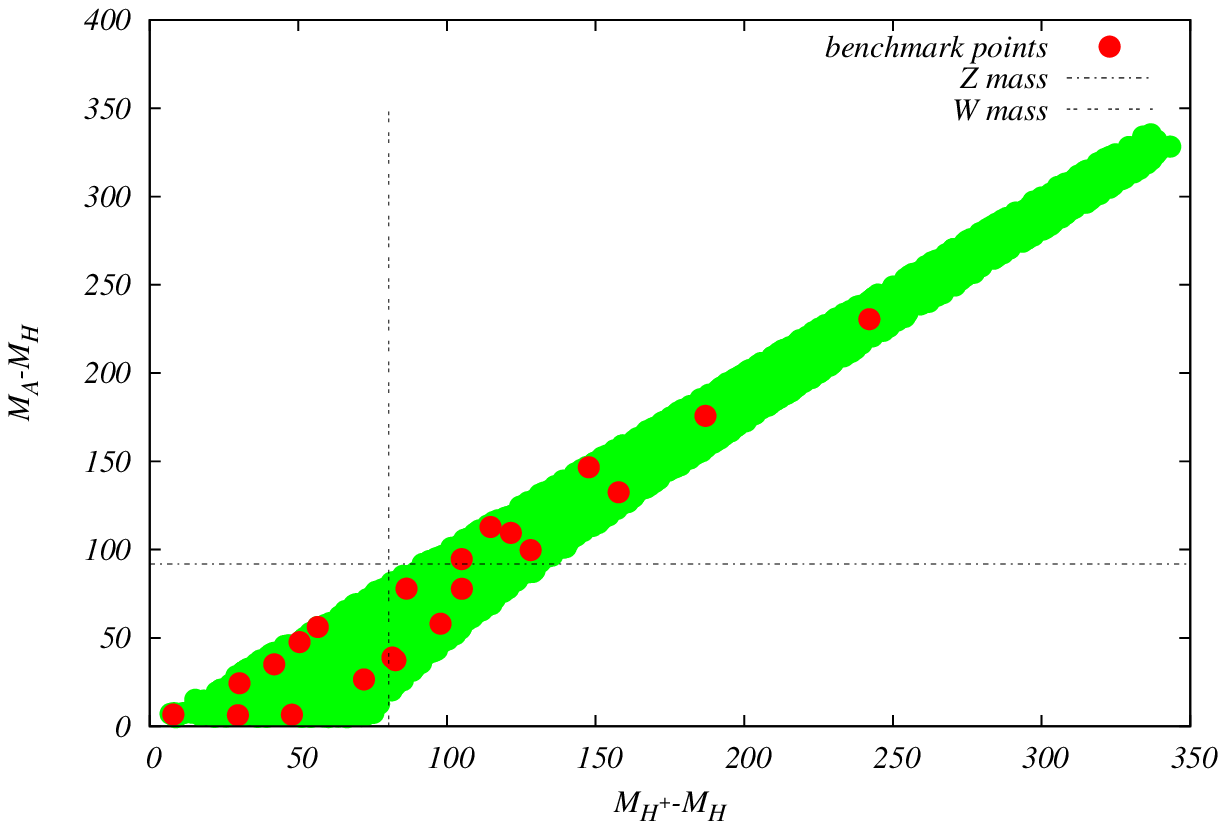}
\end{center}
\end{minipage}
\begin{minipage}{0.45\textwidth}
\begin{center}
\includegraphics[width=0.9\textwidth]{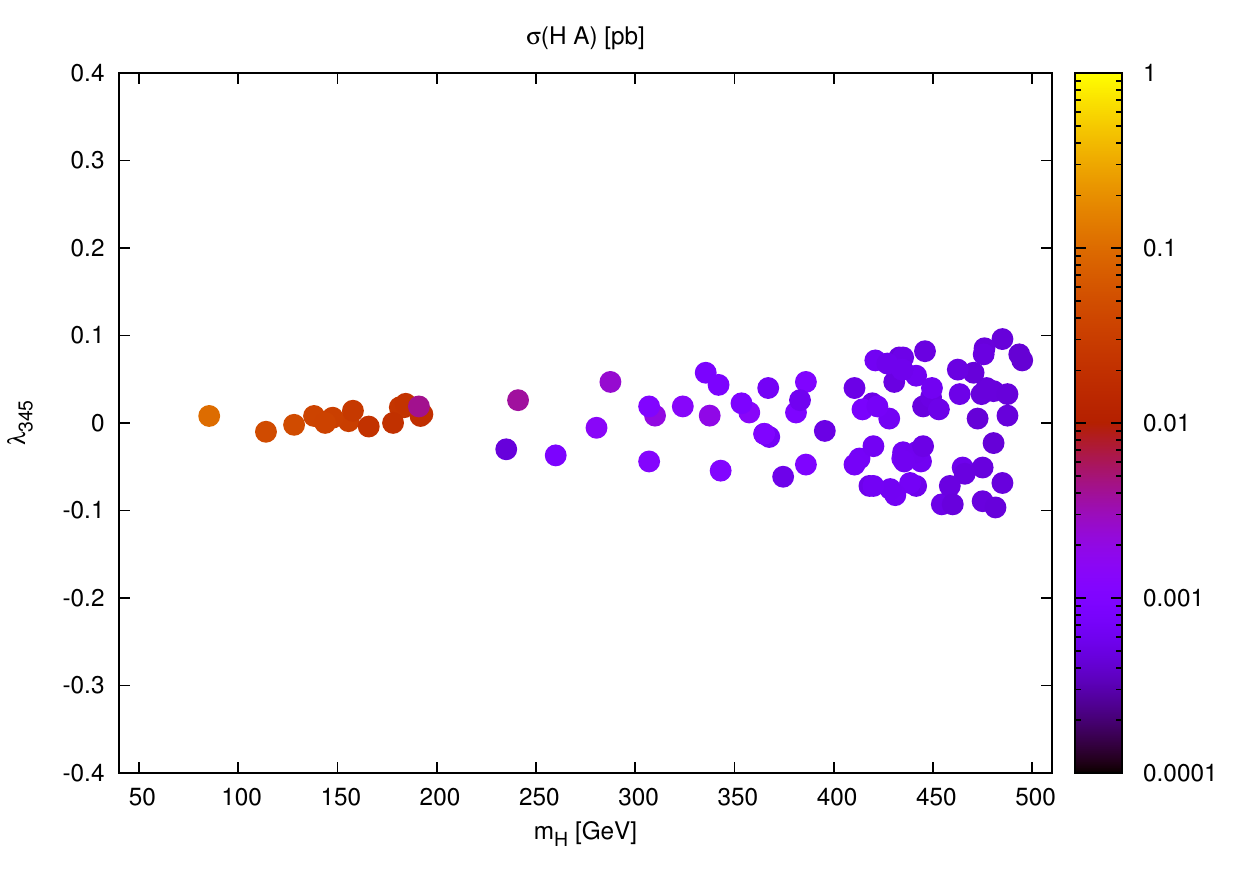}
\end{center}
\end{minipage}
\caption{\label{fig:highm} Allowed regions in parameter space after all constraints are taken into account. {\sl Left:} in the $m_{H^\pm}-m_H; m_A-m_H$ plane, taken from \cite{Kalinowski:2018ylg}. Also shown are benchmark points discussed in that work. {\sl Right:} Allowed regions in the $m_H;\lam_{345}$ plane. This corresponds to an update of figure 7 from \cite{Ilnicka:2018def}. Color-coding corresponds to the value of the leading-order production cross-sections for $HA$-production in \pb~at a 13 \TeV~LHC. These values were obtained using \mgfive{} \cite{Alwall:2011uj} with the UFO interface from \cite{Goudelis:2013uca}.}
\end{center}
\end{figure}
\end{center}

The above constraints limit the allowed regions of parameter space via an involved interplay. In particular, they limit regions for different dark matter masses, depending on the kinematic availability of the invisible on-shell decay $h\,\rightarrow\,H\,H$.
%\begin{itemize}
%\item{}
For masses $m_H\,\geq\,62.5\,\GeV$, major constraints stem from an interplay of electroweak constraints as well as direct detection limits. The former mainly determine the maximally allowed mass splitting between the dark scalars, while the latter set an upper limit on the maximally allowed value of $\lam_{345}$. We exemplarily show this in figure \ref{fig:highm}, where we display allowed points in the plane of scalar mass differences as well as in the $m_H;\lam_{345}$ plane.
%\item{}
For parameter points where $m_H\,\leq\, 62.5\,\GeV$, we refer the reader to \cite{Ilnicka:2015jba,Belyaev:2016lok,Ilnicka:2018def,Kalinowski:2018ylg} for a more detailed discussion.
%\end{itemize}

\section{IDM at colliders}

\begin{center}
\begin{figure}
\begin{center}
\begin{minipage}{0.45\textwidth}
\begin{center}
\includegraphics[width=0.8\textwidth]{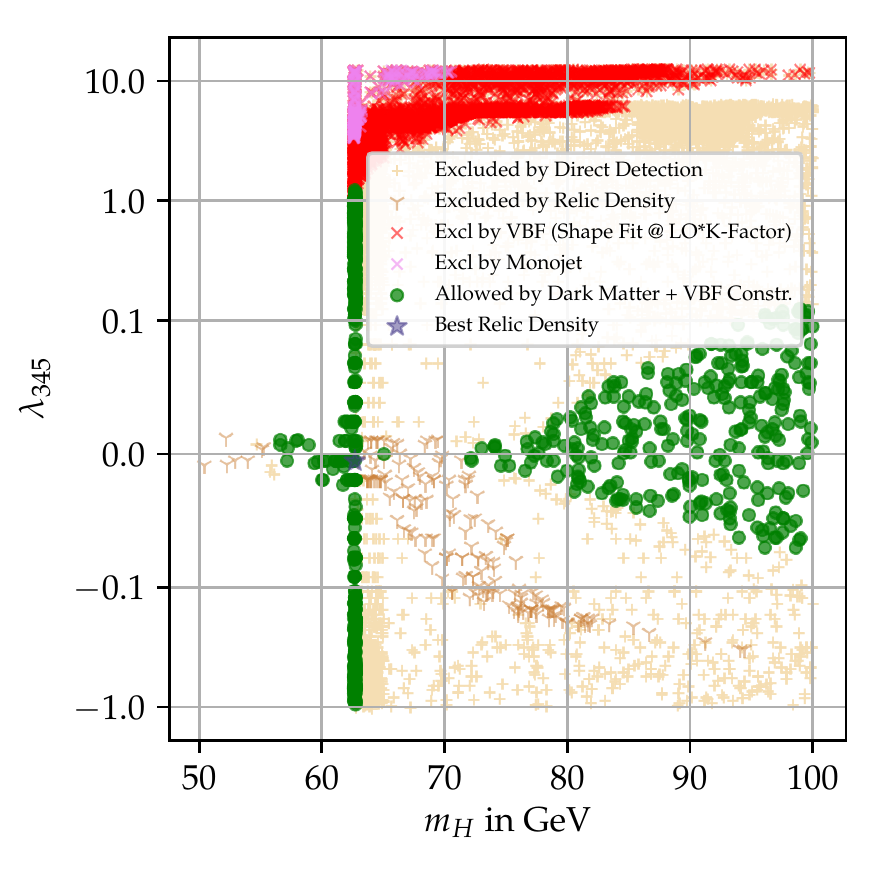}
\end{center}
\end{minipage}
\begin{minipage}{0.45\textwidth}
\begin{center}
\includegraphics[width=0.8\textwidth]{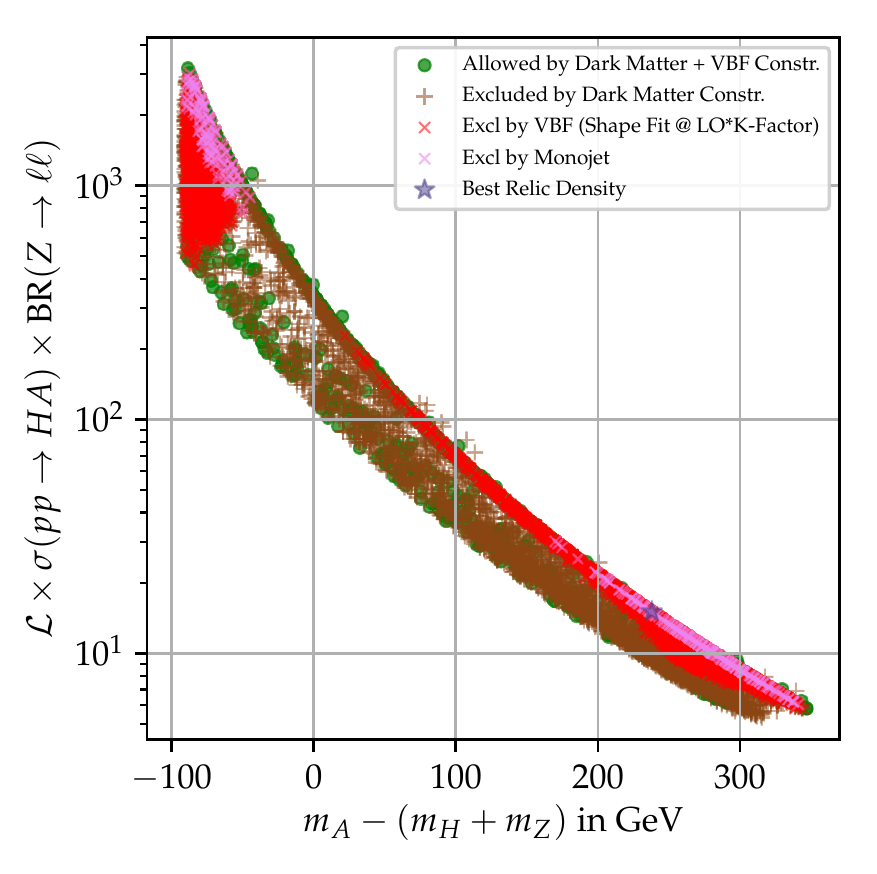}
\end{center}
\end{minipage}
\caption{\label{fig:lhcrec} Recast results for 13 \TeV searches for invisible Higgs decays, monojet, and multilepton and missing transverse energy searches \cite{Sirunyan:2018owy,ATLAS-CONF-2017-060,Aaboud:2017bja,ATLAS-CONF-2017-039}, taken from \cite{Dercks:2018wch}. {\sl Left:} parameter space in the $m_H;\lam_{345}$ plane constrained by dark matter and recast bounds. {\sl Right:} Estimated number of events in the dilepton channel as a function of kinematically available missing transverse energy. In \cite{Aaboud:2017bja}, a cut on $\slashed{E}_\perp\,\geq\,90\,\GeV$ is set, leading to a relatively small number of events even prior to additional selection cuts.}
\end{center}
\end{figure}
\end{center}

\paragraph{Hadron colliders:}
The main production channels at hadron colliders correspond to the Drell-Yan production of an $HA$ or $H\,H^\pm$ pair, followed by the predominant decay chains $A\,\rightarrow\,Z\,H;\;H^\pm\,\rightarrow\,W^\pm H$ and leading to a signature of gauge boson(s) and missing transverse energy, with on- or off-shell electroweak gauge bosons depending on the kinematic configuration and available phase space. Production cross sections for the above processes can reach up to 1 \pb~ (3 \pb) at a 13 \TeV~ (27 \TeV) LHC \cite{Ilnicka:2015jba,Belyaev:2016lok,trtalk}.
Phenomenological studies of the IDM at hadron colliders have been presented in e.g. \cite{Dolle:2009ft,Gustafsson:2012aj,Arhrib:2013ela,Belanger:2015kga,Blinov:2015qva,Ilnicka:2015jba,Poulose:2016lvz,Datta:2016nfz,Kanemura:2016sos,Akeroyd:2016ymd,Wan:2018eaz,Ilnicka:2018def,Belyaev:2018ext}. However, although many searches, e.g. within  simplified models or supersymmetric context that can lead to similar final states, have been performed by the LHC experiments, no dedicated search for this model exists. In \cite{Dercks:2018wch}, recasts of a large number of 13 \TeV~searches have been considered using the \checkmate{} \cite{Drees:2013wra,Dercks:2016npn} framework. The most constraining search for invisible decays of the SM Higgs in vector boson fusion \cite{Sirunyan:2018owy} allows to put limits on the IDM parameter space in a region with a resonantly enhanced dark matter annihilation cross section in the region $m_H\,\sim\,62.5\, \GeV$, followed by monojet searches \cite{ATLAS-CONF-2017-060}. Searches with multilepton and missing transverse energy \cite{Aaboud:2017bja,ATLAS-CONF-2017-039} were not able to pose further constraints. This is due to the fact that they usually require a relatively large cut on missing transverse energy, requiring in turn a rather large mass splitting in the dark sector that leads to smaller production cross sections. Our corresponding findings are summarized in figure \ref{fig:lhcrec}. We encourage the LHC experimental collaborations to enhance their searches for multilepton final states and missing tranvserse energy by going to lower $\slashed{E}_\perp$ cuts (see also \cite{wg3summ18}). 

\begin{center}
\begin{figure}
\begin{center}
\begin{minipage}{0.45\textwidth}
\begin{center}
\includegraphics[width=\textwidth]{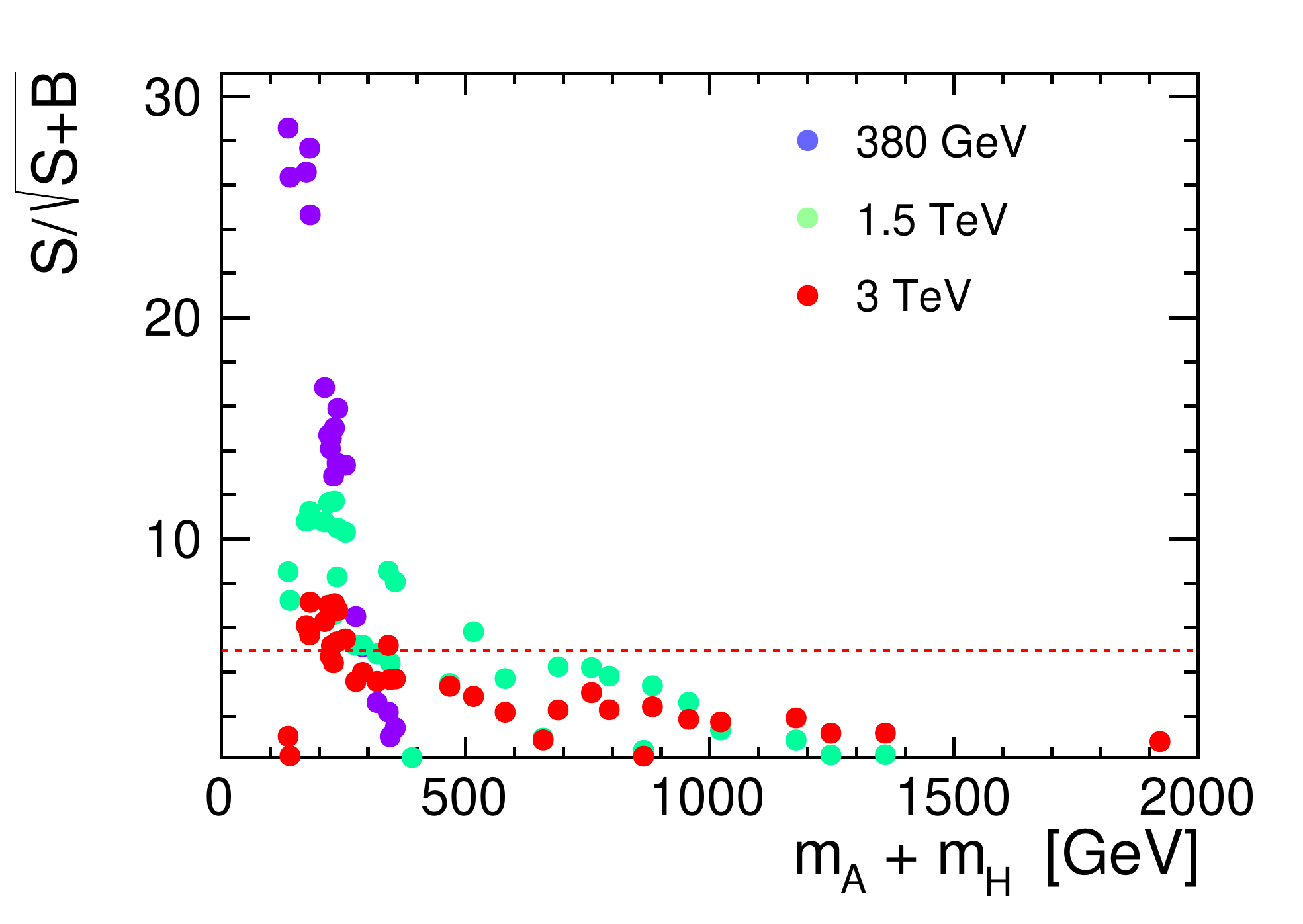}
\end{center}
\end{minipage}
\begin{minipage}{0.45\textwidth}
\begin{center}
\includegraphics[width=\textwidth]{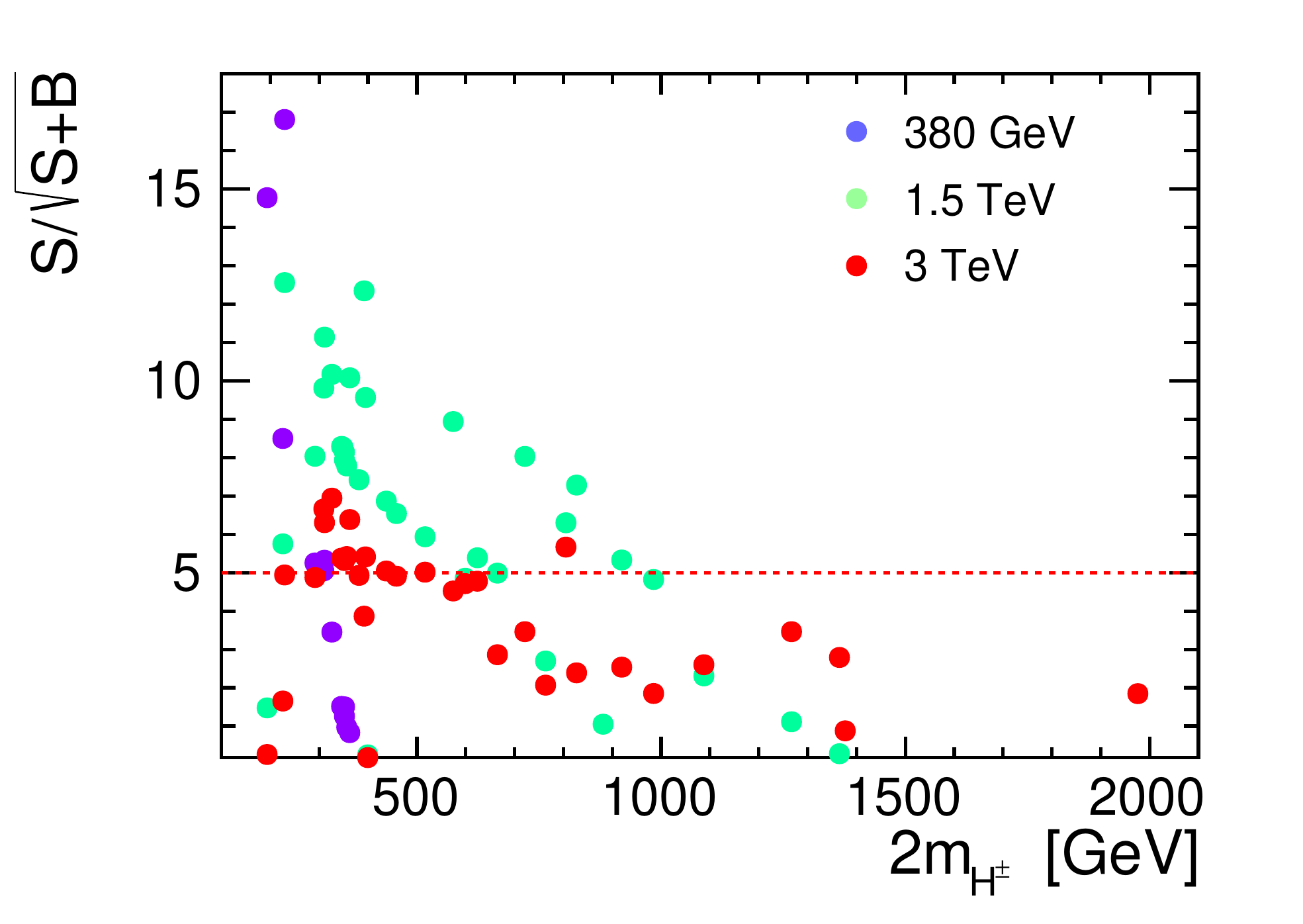}
\end{center}
\end{minipage}
\caption{\label{fig:e+e-} Projected significances, as a function of the relevant masses, for various benchmark points after full simulation and BDT analysis for various center-of-mass energies, taken from \cite{Kalinowski:2018kdn}. {\sl Left:} For dimuon and missing transverse energy ($HA$ channel). {\sl Right:} For different flavour dilepton final states ($H^+ H^-$ channel).}
\end{center}
\end{figure}
\end{center}

\paragraph{Lepton colliders:}
Benchmarks for {the Inert Doublet Model} that are in compliance with all recent bounds {and can be searched for at future lepton colliders} have been presented in \cite{Kalinowski:2018ylg,Kalinowski:2018kdn}, with results presented in \cite{Kalinowski:2018kdn,deBlas:2018mhx}.
We here focus on the main production channels
%\begin{eqnarray}
$(i)\, e^+ e^-\,\rightarrow\,H\,A;\;
(ii)\,e^+ e^-\,\rightarrow\,H^+ H^-$
%\end{eqnarray}
and subsequent leptonic decays into $(i)$ dimuon and $(ii)$ different flavour dilepton final states accompanied by missing transverse energy.
Production cross sections for the above channels can reach up to $\sim\,150\,\fb$ for collider scenarios with center-of-mass energies {in the 250--500\,GeV range}~\cite{Kalinowski:2018ylg}. In \cite{Kalinowski:2018kdn}, a detailed analysis was presented for the potential {of} CLIC \cite{Linssen:2012hp,Aicheler:1500095,CLIC:2016zwp,Robson:2018zje} {running at} center-of-mass energies of 380 \GeV~{with} an integrated luminosity of $1\,\ab^{-1}$ as well as 1.5 and 3 \TeV~ {with} $2.5\,\ab^{-1}$ and $5\,\ab^{-1}$, respectively \cite{Robson:2018zje}. Signal and background were simulated using {\tt WHizard 2.2.8} \cite{Moretti:2001zz,Kilian:2007gr}. For the signal, we made use of the corresponding SARAH \cite{Staub:2015kfa}/ {\tt SPheno 4.0.3} \cite{Porod:2003um,Porod:2011nf} interface.
In the simulation, we did not specify intermediate states but instead considered all processes leading to dilepton final states and missing transverse energies, including processes with up to four additional neutrinos. After preselection cuts, a boosted decision tree (BDT) algorithm as implemented in the TMVA toolkit \cite{Hocker:2007ht} was employed to optimize selection criteria. Beam spectra and beamstrahlung were included in all analyses. Our findings are summarized in figure \ref{fig:e+e-}, where we plot the expected significance as a function of the relevant mass scales. We find that at a center-of-mass energy of 380~\GeV, scales up to 300 \GeV~ are accessible. An increase in collider energy enhances this reach up to 550 \GeV~for the dimuon and 1 \TeV~ for the different flavour final states.

\section{Conclusions}\label{sec:conclusions}
We discussed the current status of the Inert Doublet Model, a two Higgs doublet model with a discrete symmetry that features a dark matter candidate. The model is already relatively strongly constrained, with major limits stemming from electroweak precision data, theoretical bounds on the potential and the couplings, as well as dark matter findings. We discussed limits and discovery prospects at current and future hadron colliders as well as $e^+ e^-$ machines.

\section{References}
%\printbibliography
%\bibliographystyle{h-elsevier}
%\bibliographystyle{iopart-num}
%\bibliography{lit}
\input{dis_procs-3.bbl}

\ack{
We acknowledge funding by the National Science Centre,
Poland, the HARMONIA project under contract UMO-2015/18/M/ST2/00518
(2016-2019) and OPUS project under contract UMO-2017/25/B/ST2/00496
(2018-2021), COST Action CA 15180, the National Science Foundation under
Grant No. 1519045, by Michigan State University through computational resources
provided by the Institute for Cyber-Enabled Research, by grant K 125105 of
the National Research, Development and Innovation Fund in Hungary, the German Research Foundation (DFG) under grants
number STO 876/4-1 and STO 876/2-2, and the SFB~676 ``Particles, Strings and the Early Universe''.} 

\end{document}

%% file: dis_procs-3.bbl
\providecommand{\newblock}{}